\documentclass[preprint2,longabstract]{aastex}

 \shorttitle{Heavy Ion
Heating} \shortauthors{Korreck, Zurbuchen, Lepri, \& Raines}
\begin{document}
\title{Heating of Heavy Ions by Interplanetary Coronal Mass Ejection
(ICME) Driven Collisionless Shocks}

\author{K. E. Korreck\altaffilmark{1,2},
T. H. Zurbuchen\altaffilmark{2}, S. T. Lepri\altaffilmark{2}, \&
J. M. Raines\altaffilmark{2} }

\altaffiltext{1}{Harvard Smithsonian Center for Astrophysics, 60
Garden Street, Cambridge, MA 01238}

\altaffiltext{2}{University of Michigan, Atmospheric, Oceanic, and
Space Science Department, Ann Arbor, MI 48109}

\begin{abstract}
Shock heating and particle acceleration processes are some of the most
fundamental physical phenomena of plasma physics with countless
applications in laboratory physics, space physics, and
astrophysics. This study is motivated by previous observations of
non-thermal heating of heavy ions in astrophysical shocks
\citep{kor04}. Here, we focus on shocks driven by Interplanetary
Coronal Mass Ejections (ICMEs) which heat the solar wind and
accelerate particles. This study focuses specifically on the heating
of heavy ions caused by these shocks. Previous studies have focused
only on the two dynamically dominant species, H$^+$ and
He$^{2+}$. This study utilizes thermal properties measured by the
Solar Wind Ion Composition Spectrometer (SWICS) aboard the Advanced
Composition Explorer (ACE) spacecraft to examine heavy ion heating. This
instrument provides data for many heavy ions not previously available
for detailed study, such as Oxygen (O$^{6+}$, O$^{7+}$), Carbon
(C$^{5+}$, C$^{6+}$), and Iron (Fe$^{10+}$).  The ion heating is found
to depend critically on the upstream plasma $\beta$, mass per charge
ratio of the ion, M/Q, and shock magnetic angle,
$\theta$$_{Bn}$. Similar to past studies \citep{sch88}, there is no
strong dependence of ion heating on Mach number. The heating mechanism
described in \citet{lee00} is examined to explain the observed heating
trends in the heavy ion thermal data.

\end{abstract}


\section{Introduction}
Shocks constitute a non-linear transition between two dynamically
different plasma states. At this transition, a portion of the bulk
flow energy of the plasma is transferred into thermal energy of its components. A very small fraction (1-10\%) of the interacting plasma
becomes highly energized and is injected into particle acceleration
processes boosting the particles to very high energy (E$\ge$
GeV). There is, however, a lack of understanding of the dynamic
effects that lead to this heating.  The heating of an ion species
appears to be dependent on the ion properties as well as on the bulk
properties of the shock and the initial state of the plasma
\citep{shi,lee00,kor04,ber97}.  It is the purpose of this paper to
investigate the heating of solar wind ions using accessible thermal
data of heavy ions from shocks driven by the interplanetary
manifestations of Coronal Mass Ejections (CMEs) - Interplanetary CMEs
(ICMEs).

\citet{lee00} discuss three parameters key to understanding heating in
collisionless shocks: Alfvenic Mach number, M$_{A}$, magnetic shock
angle, $\theta$$_{Bn}$, and upstream plasma $\beta$, the ratio of
thermal to magnetic energy.  The dependence of heating on these
parameters has been examined before in solar wind shocks both
observationally and theoretically \citep{ogi80, zer76, ber97, zha91}.
Work by \citet{ber97} examined the heating of ions in the
solar wind.  The heating of heavy ions such as O$^{6+}$ was found to
be more than mass proportional when compared to the heating of
protons.  This appears to contradict heating seen in other shocks such
as those around supernova remnants \citep{kor04}.  The \citet{ber97}
paper also found the ion heating to be most efficient when the initial
ion and proton thermal velocity were similar, i.e. energy is distributed mass proportionally.

The effect of Alfvenic Mach number of the shock on heating has been
studied observationally in the heliosphere.  The Alfvenic Mach number
is used instead of the acoustic Mach number due to the magnetic nature
of the solar wind.  \citet{sch88} used data from crossings of the
Earth's bow shock to study the heating of ions and electrons.  Only a
weak correlation between increasing Mach number and decreased heating
was found.

The plasma $\beta$ is a defining characteristic of the plasma.  For
example, \citet{lip99} found for plasmas with $\beta$ $\ge$ 0.1 ions were unable to ``surf'' waves as an acceleration process, making plasma $\beta$ a determining factor in particle acceleration
models.  Note that in the heliosphere $\beta$ is normally $\le$ 1.0
\citep{gal97}.

The magnetic angle from the shock normal, $\theta_{Bn}$, has also been
examined with respect to the acceleration of ions.  Quasi-parallel
shocks have been shown to accelerate ions from thermal energies to
well above what would be expected from the bulk energy transfer
\citep{ell81, sch90, gia92, kuc95}.  Quasi-perpendicular shocks have
been shown to be inefficient for accelerating ions from thermal ion
populations, however these shocks are the most efficient accelerators
if there is a seed population available.  The acceleration efficiency,
once the seed population is available, is inversely proportional to
$\theta$$_{Bn}$ \citep{ell95}; the more perpendicular the shock, the
more efficient the acceleration.  Hence, in order to understand cosmic
ray acceleration, the heating of a suprathermal seed particle population is
of great importance.

In addition to the three parameters mentioned above, the
mass-to-charge ratio also plays a crucial role in heating of heavy
ions.  As the mass-to-charge ratio increases the heating of ions
increases \citep{lee00}.  The mass-to-charge ratio dictates which
types of MHD wave interactions are available to the particles at the
shock front.

The novel data on ICMEs from the Advanced Composition Explorer (ACE)
spacecraft \citep{sto98} afford us the opportunity to study  these
shocks and the heating processes for individual ion species in greater
detail.  While the previous work included only protons, helium, and
oxygen (O$^{6+}$) in the analysis, this study is extended to include
O$^{7+}$, C$^{5+}$, C$^{6+}$, and Fe$^{10+}$ ions, extending the
mass-to-charge ratio range to 1 thru 5.6.

This paper examines the heating of heavy ions with respect to all
critical parameters mentioned above that dominate shock heating - the
magnetic shock geometry - $\theta$$_{Bn}$, Alfvenic Mach number -
M$_{A}$, plasma $\beta$, and $M/Q$, the mass per charge ratio.  We
utilize thermal properties measured by the Solar Wind Ion Composition
Spectrometer (SWICS) \citep{glo98} on the ACE spacecraft. We
characterize a statistically significant sample of 20 shocks for which
heavy ion heating is observed with high statistical accuracy and with
well-defined shock properties. Magnetic field data from the
Magnetometer (MAG) instrument \citep{smi98} and proton data from the
Solar Wind Electron Proton Alpha Monitor (SWEPAM) instrument
\citep{mcc98} were also used.

The plasma data available from the ACE spacecraft and a brief
discussion of errors are described in Section 2.   The selection of
ICME shocks is discussed in Section 3. In Section 4, the heating of
ions is examined with respect to each parameter.  A method of heating
is described in Section 5 and conclusions are summarized in Section 6.

\section{Observations}

The ACE spacecraft, launched in 1997, orbits about the first
Lagrangian, L1 point, 235 Earth radii upstream along the Sun-Earth
line.  Three of the nine instruments aboard were used for this study.
The SWICS, MAG, and SWEPAM instruments provide data on ion
composition, velocity, density, and the ambient magnetic field.   For
this study, the following physical quantities were analyzed: proton
temperature, proton thermal velocity, proton number density, thermal
velocities of He$^{2+}$, O$^{6+}$, O$^{7+}$, C$^{5+}$, C$^{6+}$, and
Fe$^{10+}$, and magnetic field.

The heavy ion data (Z$>$1) was obtained from the SWICS instrument.
This instrument is composed of an electrostatic analyzer, which
measures an ion's energy per charge, and a time-of-flight-energy
telescope, which measures an ion's velocity and total energy. The
combination of these two sensors enables unique identification of a
particle's mass, charge, and energy \citep{glo98}.  For this study, we
use SWICS data with 1-hour time resolution; this time resolution
provides sufficient statistics for the minor ions studied.  For each
one hour time bin, five 12-minute accumulation cycles were
summed. Care has been taken to exclude the accumulation cycle that
contains the moment of the shock passage, because it mixes upstream
and downstream distributions. The proton thermal velocity, the proton
number density, and the magnetic field data were taken from the
combined SWEPAM/MAG data set available on the web
at:\\http://www.srl.caltech.edu/ACE/ASC/. 64-second averages of data
were used.  The data from each physical quantity were averaged for one
hour before or one hour after the shock.  This allows for a relatively
local measurement of the shock, but statistically significant heavy
ion data. There are important statistical limits to these data, which
are discussed in section \ref{errors}.

\subsection{Errors\label{errors}}

For the proposed heating analysis, it is necessary to quantify the
thermal width, or thermal speed, of the heavy ion
distributions. However, the thermal speed is only defined if
distributions are Gaussian. We have therefore implemented a test of the fit of a Gaussian to the data using the following technique.  For each heavy ion measurement,
the thermal speed was calculated using two techniques: by calculating
the second moment of the observed distribution, $v_{th}$, and by
determining the thermal speed by least-squares fit of a Gaussian to
the distribution, $ v_{th,fit}$.  The deviations between these two
quantities are a measure of the deviations of the distributions from a
Gaussian shape.

The deviations were relatively small for most ions with an average
value of $<$10\%, but we do find a few examples for which energetic
tails on the distribution function dominate, and the error on $
v_{th}$ was then increased, to a more conservative error estimate of
20\% in order to account for this.

In order to combine plasma data with our composition-resolved dynamics
data, we averaged the necessary solar wind parameters for 1 hour ahead
of and behind the shock front using data taken at intervals of 64
seconds, to overlap with the SWICS data-interval.  The statistical
error was then calculated as the standard deviation of each averaged quantity.

\section{Shock Selection} 

Shocks were selected from the \citet{cra03} list of ICMEs and the ACE
Shock List maintained online
\footnote{\\\url{http://www.bartol.udel.edu/$\sim$chuck/ace/ACElists/obs\_list.html}}.
The ACE Shock List details the time of the shock passage, the angle
between upstream magnetic field vector and the shock normal,
$\theta_{Bn}$, and the upstream Mach number, M$_{A}$

The first criterion for selecting a shock for this study was to have
all crucial data-sets available for two hours before and two hours
after the shock passage.  The next criterion for selection was the
characteristics of the kinetic temperature of the solar wind before
the shock passage.  Since the kinetic temperature of the solar wind
provides a measure of the shock and the plasma characteristics
upstream and downstream of the shock, an increase in temperature
before the time of the shock indicates pre-heating, ion feedback or a
reverse shock \citep{pas81, glo99}.  The presence of these features
would bias the study since ions are heated differently, hence these
instances need to be excluded. If the temperature increased in the
time period between 60 to 30 minutes prior to the shock passage by
more than 30\% of the mean value of the temperature calculated up to
30 minutes before the shock, the shock was considered to exhibit
pre-heating and was excluded from the analysis.

\begin{figure}
\begin{center}
\includegraphics[angle=90,clip=,scale=1]{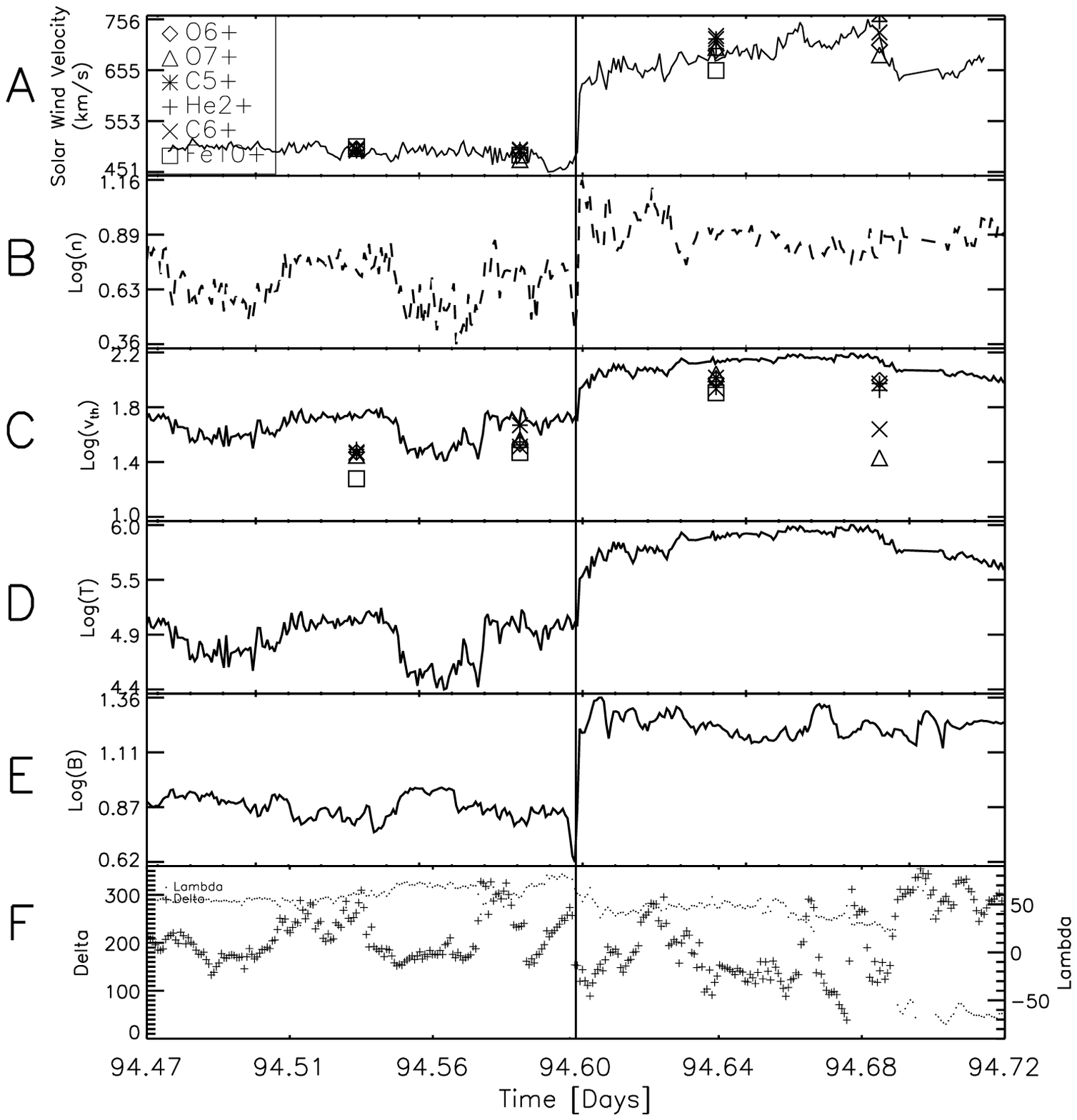}
\caption{Plot of ACE magnetic and temperature data versus time, in
fraction of a day, for a parallel shock.  Panel A plots the solar wind
proton velocity, v$_{p}$ as the solid line and the velocity of each
heavy ion is included as a different symbol.  Panel B is plot of the
proton number density, n.  Panel C is a plot of the thermal velocity,
v$_{th}$, of protons with the symbols representing the thermal
velocity of individual heavy ions. Panel D plots the proton
temperature, T, versus time.  Panel E contains the magnitude of the
magnetic field, B. Panel F is a plot of the magnetic latitude,
$\delta$, and longitude, $\lambda$.\label{goodpar}}
\end{center}
\end{figure}

Representative solar wind parameters for an ICME shock are plotted
versus time in Figure \ref{goodpar}.  Panel A shows the solar wind
proton velocity, $v_{p}$, versus time as the solid line and the
velocity of each heavy ion is included as a different symbol, as
indicated in the plot legend.  Panel B shows the proton number
density, n, in the solar wind. Panel C is a plot of the thermal
velocity, $v_{th}$, of protons with the over-plotted symbols
representing the thermal velocity of individual ions. Panel D plots
the solar wind proton temperature, $T_{p}$, versus time. Panel E
contains the magnitude of the magnetic field, B, versus time.  Panel F
is a plot of the magnetic latitude, $\delta$, and longitude,
$\lambda$, versus time.

In order to examine the dependence of heating on shock orientation
with respect to the background magnetic field, the shock list was
broken up into quasi-parallel and quasi-perpendicular
shocks. Quasi-parallel shocks were defined to have  $\theta_{Bn}$
between 0 and 20 degrees and shocks with $\theta_{Bn}$ between 80 and
90 degrees were considered quasi-perpendicular.  There were 16
quasi-perpendicular shocks available for study and 4 quasi-parallel
shocks.  In subsequent sections, each parameter will be explored for a
perpendicular and parallel shock separately.

\section{Heavy Ion Heating}

There are two methods of heating ions that are widely invoked in the
study of collisionless shocks.  First, the transfer of the bulk fluid
kinetic energy to the thermal energy of the particles leading to mass
proportional heating or all ions having the same velocity.  The second
method is wave particle interactions that would result in a M/Q
dependence in heating.

Another point of interest is the distribution of heating among the
different ion species.  In studies in the heliosphere \citep{ber97},
it was found that ions are heated more than mass proportionally than
ions.  However, in studies of the collisionless shocks in supernova
remnants, the ions are heated less than mass proportionally to the
protons \citep{kor04}.  \citet{lee00} found that the heavy ions are heated by direct transmission through the shock as reflection of heavy ions by the shock potential is prohibitive due to their large mass.

In order to determine which mechanism or mechanisms are responsible
for the heating, we first quantify the heating of each ion species,
including protons, and then examine the heating dependence on M/Q,
M$_{A}$, and $\beta$ to try to elucidate a candidate mechanism for
heating.

In this study the heating, H, of an ion at a shock is defined as the
ratio of the square of the upstream and downstream thermal velocities.
\begin{equation}\label{heatingeq}
H=\frac{v_{th_{d}}^{2}}{v_{th_{u}}^{2}}=\frac{3kT_{s,d}/m_{s}}{3kT_{s,u}/m_{s}}=\frac{T_{s,d}}{T_{s,u}}
\end{equation}

where\\ T$_{s}$= Temperature of the species\\  m$_{s}$= Mass of the
species\\  v$_{th_{d}}$= Average thermal velocity of the species one
hour downstream\\  v$_{th_{u}}$= Average thermal velocity of the
species one hour upstream\\  k= Boltzmann constant \\

However, it has been recognized that the solar wind is not initially
in thermal equilibrium: heavy ions have a tendency to exhibit high
kinetic temperatures, \citep{hef98, zur00, von03}.  In order to
accurately describe the ion heating, the preshock temperature ratio of
ion temperature to proton temperature needs to be considered.  The
ratio of preshock ion to proton temperature shows only one of the 96
data points to be significantly less than 1.  The data in general
supports an initial condition where ions have equal if not higher
thermal speeds than the protons. Therefore, relative heating ratios based
on initial thermal speeds will give the clearest quantification of
heating through the shock passage.

Examining the current data set with respect to the temperature ratio
of the downstream ions to protons, we have found that the heating is
on average less than mass proportional for the ions except Fe$^{+10}$,
although 27 out of the 96 data points for perpendicular shocks did
show greater than mass proportional heating.  This contradicts earlier
studies by \citet{ber97} that found all of the ion heating to be
greater than mass proportional.

\begin{figure}
\includegraphics[angle=90,clip=,scale=0.4]{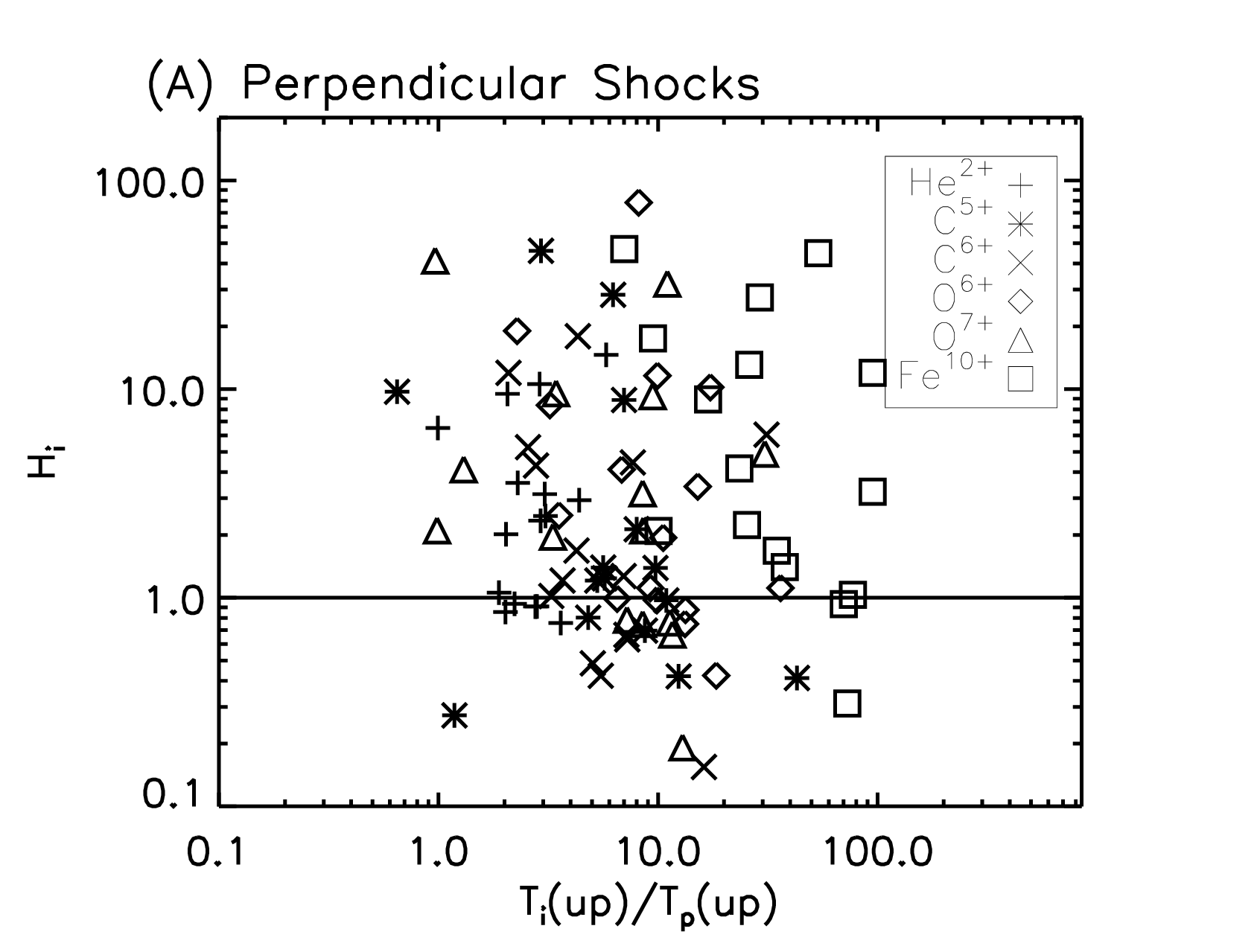}
\includegraphics[angle=90,clip=,scale=0.4]{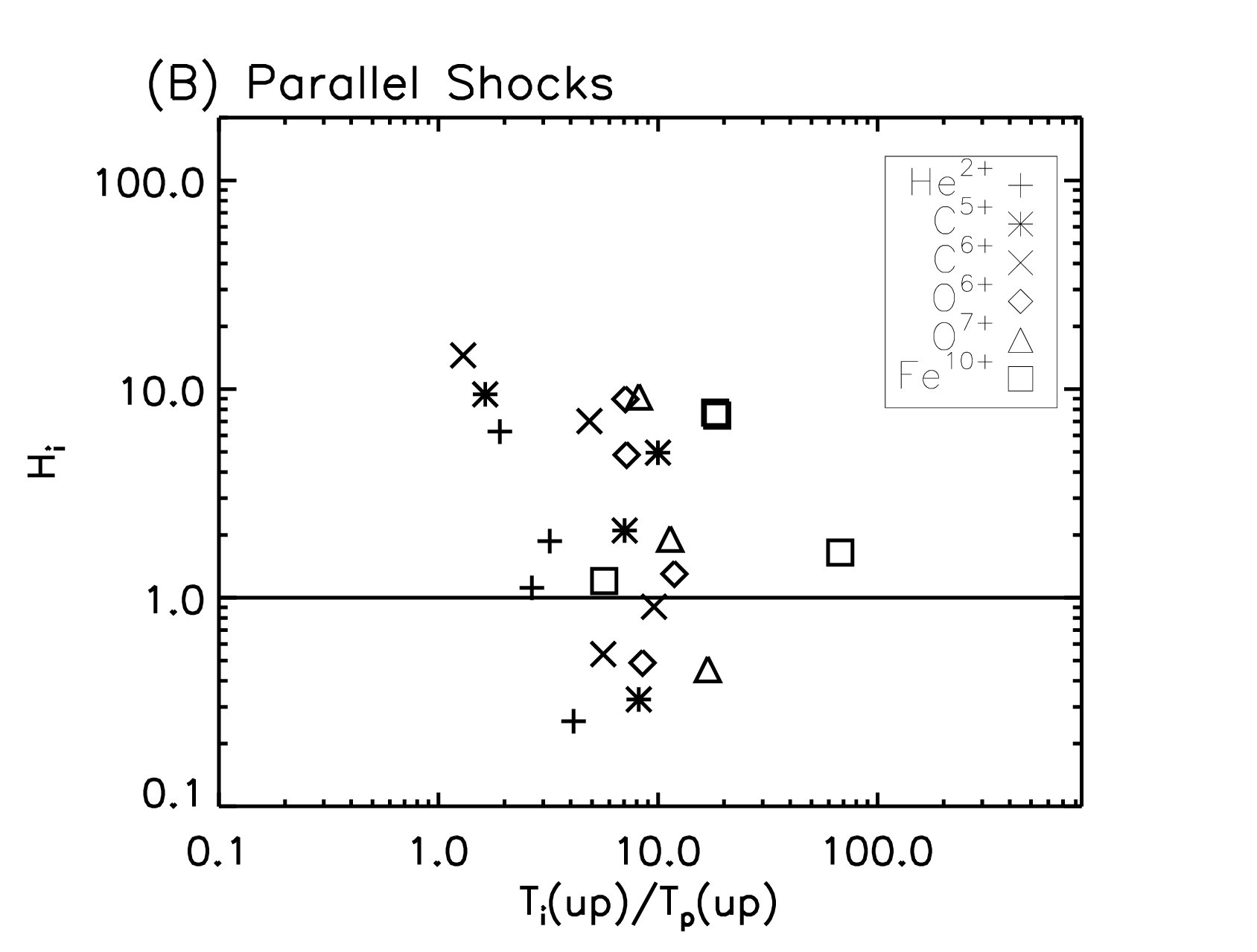}
\caption{\label{hratios}Shock ion heating versus upstream thermal
temperature ratio, (a) for perpendicular shocks and (b) for parallel
shocks.  The upstream ratio of ion thermal velocity to proton thermal
velocity, T$_{i}$/T$_{p}$, is the x-axis.  Ion heating, H$_{i}$, is
the y-axis.  }
\end{figure}

The heating of the ions are plotted versus the initial temperature
ratio of the ion to the proton in Figures \ref{hratios}a and
\ref{hratios}b.  The values of initial ratio are all greater than or
equal to one except for one data point.  The heating, if greater than
one, indicates that the species was heated through the shock.
However, only 69\% of the ions were heated in the perpendicular shocks
and 70\% in the parallel shocks.  The others actually showed a
decrease in temperature across the shock front. Protons all gained
heat or maintained their temperature across the shock front.

\begin{figure}
\includegraphics[angle=90,clip=,scale=0.4]{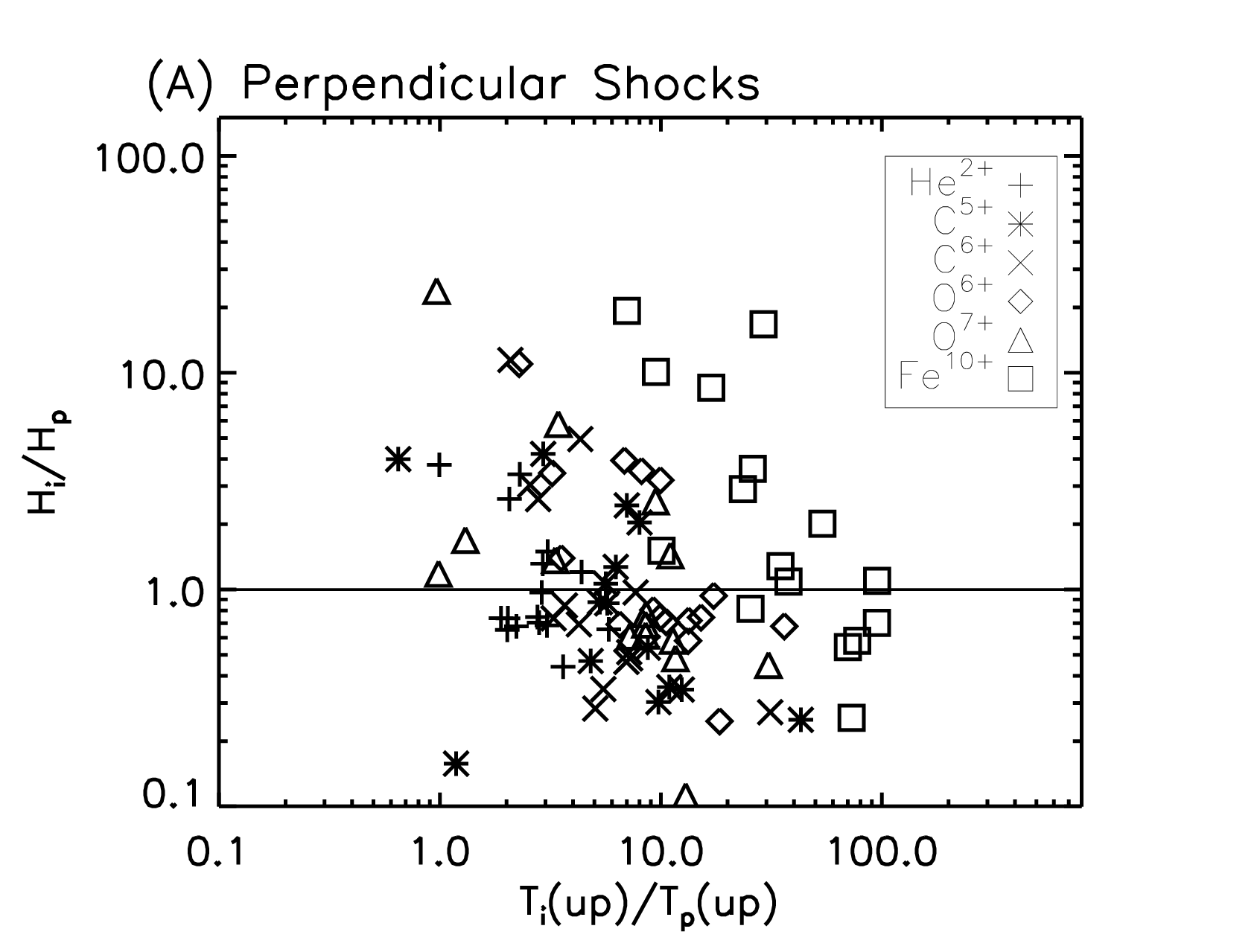}
\includegraphics[angle=90,clip=,scale=0.4]{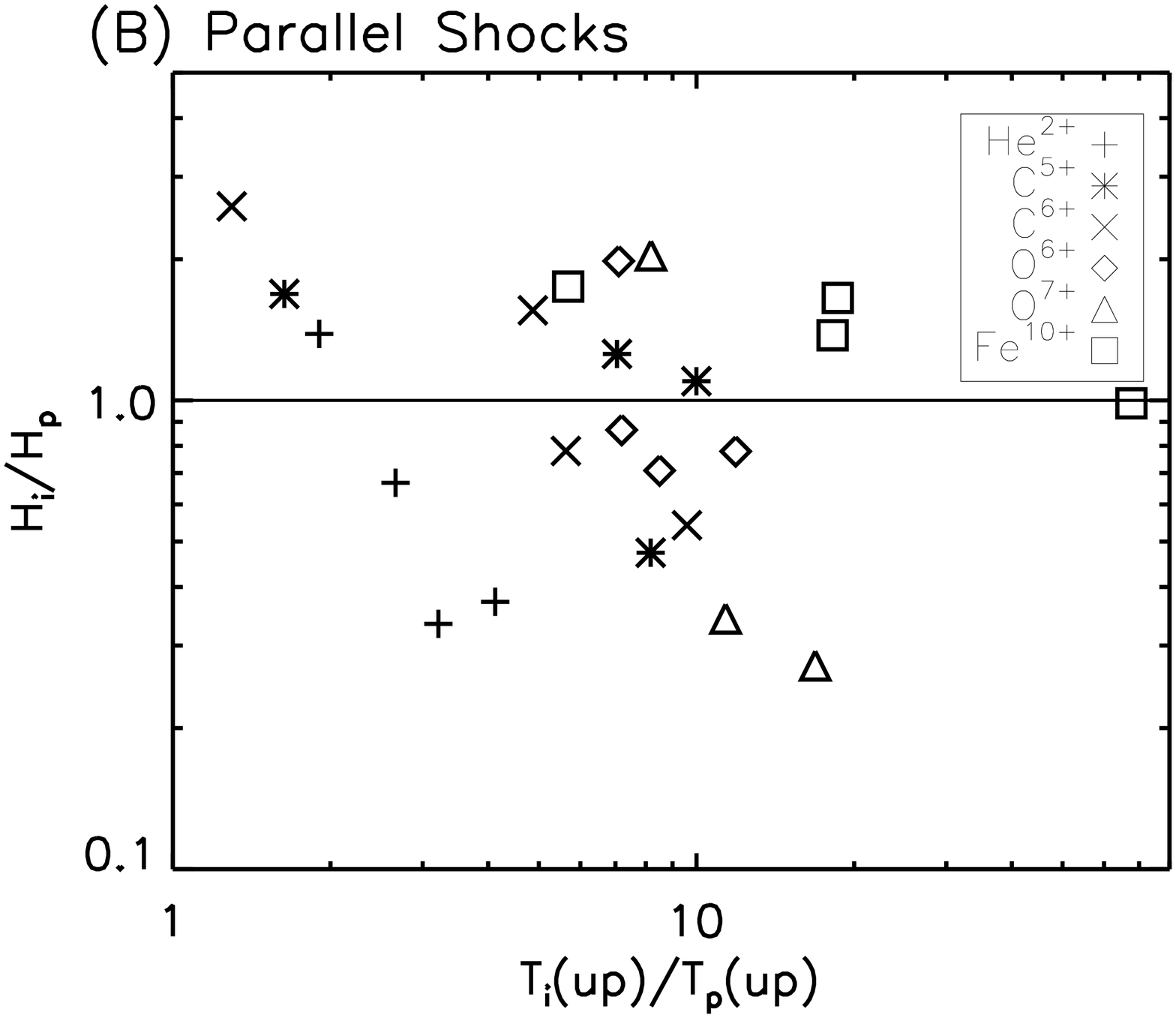}
\caption{\label{myberd}Shock heating ratios versus upstream thermal
temperature ratio, (a) for perpendicular shocks and (b) for parallel
shocks.  The upstream ratio of ion temperature to proton temperature,
T$_{i}$/T$_{p}$, is the x-axis.  The ratio of temperature increase
between the ion, H$_{i}$ is the y-axis.  }
\end{figure}

In Figures \ref{myberd}a and \ref{myberd}b, the ratio of ion and
proton heating, $H_{i}/H_{p}$, is plotted versus the initial ratio of
thermal speeds of the ion and proton where $H_{p}$ is the heating
ratio for protons.  Only 43\% of the ions in perpendicular shocks were
heated more than the protons and 50\% of those ions in parallel
shocks. The protons are therefore gaining more heat than the ions in
these shocks.

To characterize the heating mechanism consistent with the current data
set, the dependence of the heating on Alfvenic Mach number, M$_{A}$,
plasma $\beta$, and mass to charge ratio, M/Q, for both quasi-parallel
and quasi-perpendicular shocks will now be examined.

\subsection{Dependence of Heating on Alfvenic Mach Number}

 Shocks can heat ions by converting bulk kinetic energy into thermal
kinetic energy of the particles. The Mach number gauges the speed of
the propagation of bulk fluid flow relative to the surrounding medium.
The Mach number used in this study is the Alfvenic Mach number, which
takes into account interaction with the surrounding magnetized plasma.
Previous studies, \citep{sch88} showed the collisionless heating to be
weakly dependent on Mach number.

The method of energy transfer can be determined by the Critical Mach
number. Below an Alfvenic Mach number of approximately 2
\citep{ken87}, also called a sub-critical shock, the effective
dissipation of energy by the shock is local, i.e. the heating is
spatially local to the shock front via conduction.  Above this
critical Mach number the dissipation is based on multiple streams of
ions which have longer characteristic scales. A convenient measure of
a supercritical shock is the ratio of the Alfvenic to Magnetosonic
Mach numbers.  If this ratio is $\ge$ 1 then the shock is
supercritical by this definition.  We have calculated that all the
shocks in this study are supercritical.  The data therefore implies a
heating method that involves wave particle interaction.  Protons heating has been shown to result from both direct transmission and relflected protons in these supercritical shocks \citep{lee87}.

\begin{figure}
\includegraphics[angle=90,clip=,scale=0.4]{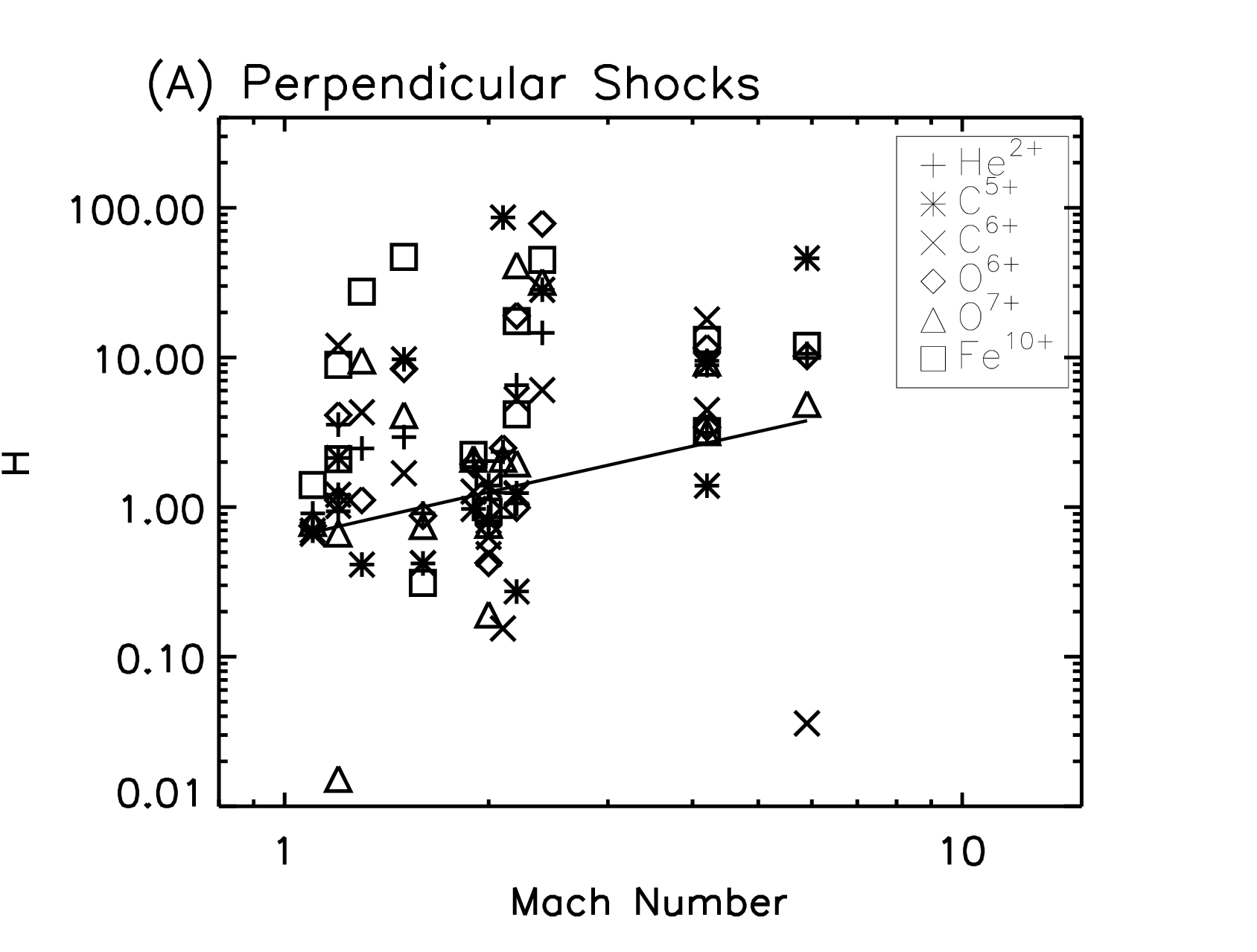}
\includegraphics[angle=90,clip=,scale=0.4]{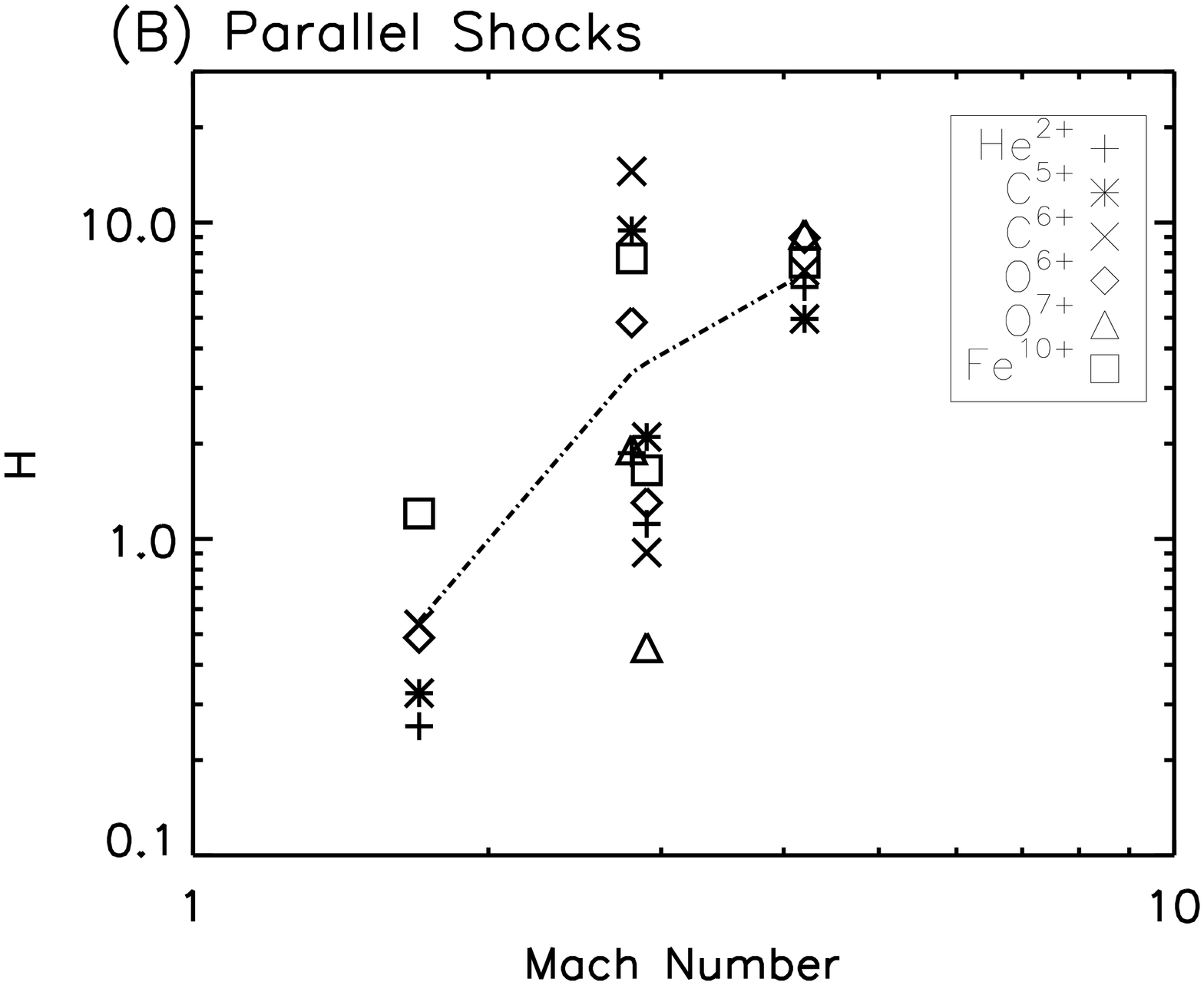}
\caption{\label{machf} Plot of Alfvenic Mach number versus heating, H,
for all the heavy ions. The line indicates a least squared fit to the data which appears kinked due to the logrithmic scale. }
\end{figure}

Figure \ref{machf}a, shows the Alfvenic Mach number versus the heating
for quasi-perpendicular shocks.  The heating, H, is described by
Equation \ref{heatingeq}.   The line shows a linear fit to the data
with an increasing trend. The parallel fit seems kinked due to the logrithmic scale for the data. The quasi-parallel shocks in Figure
\ref{machf}b also shows an increasing trend with increasing Mach
number however with a smaller data set and less statistics it seems to
be a weaker effect.  This contradicts the findings of \citet{sch88}, however the fits are marginal ($\chi_{\nu}^{2}$ $\sim$ 5 for the parallel fit and 0.6 for the perpendicular fit).

\subsection{Plasma $\beta$ Effect on Heating}

\begin{figure}
\includegraphics[angle=90,clip=,scale=0.4]{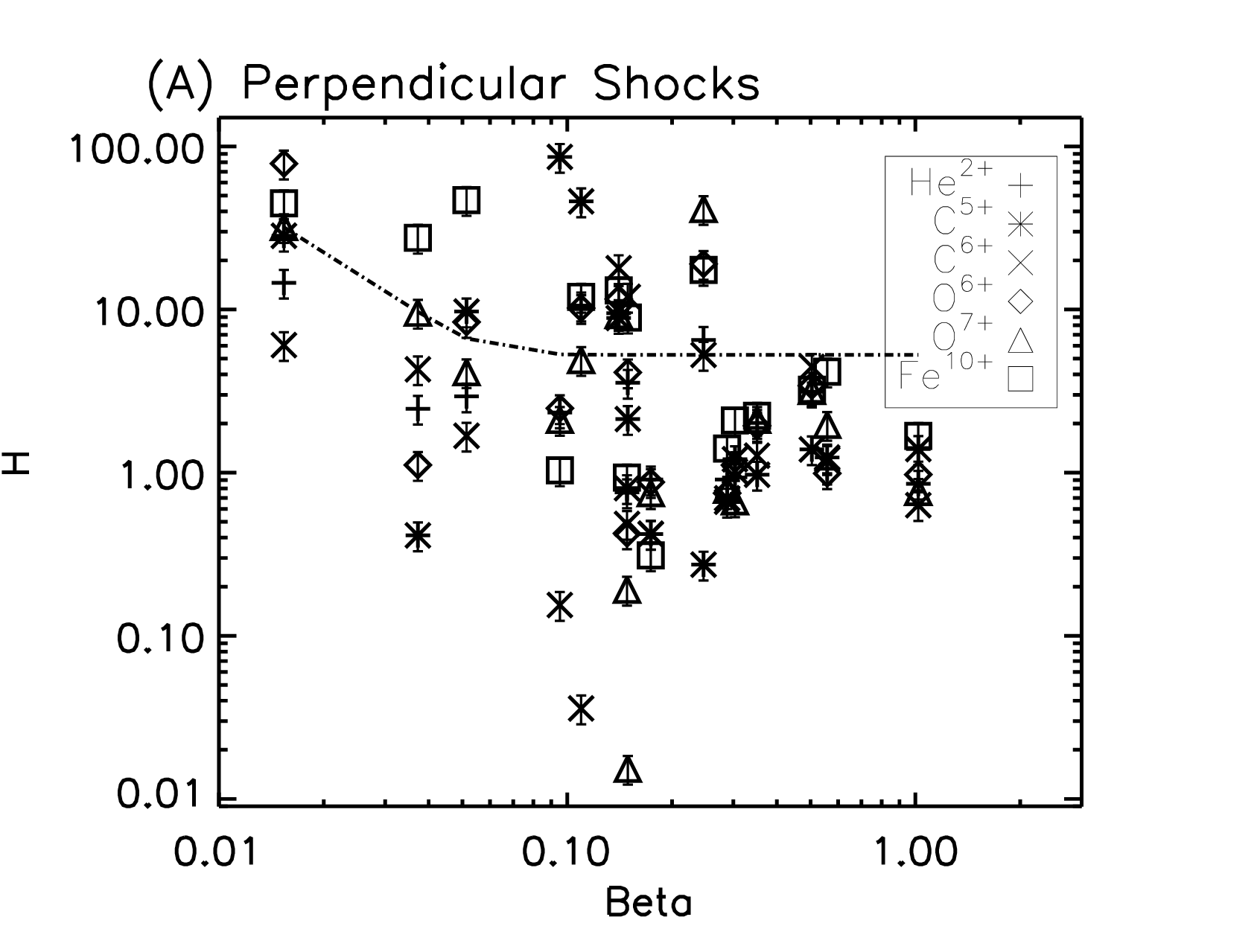}
\includegraphics[angle=90,clip=,scale=0.4]{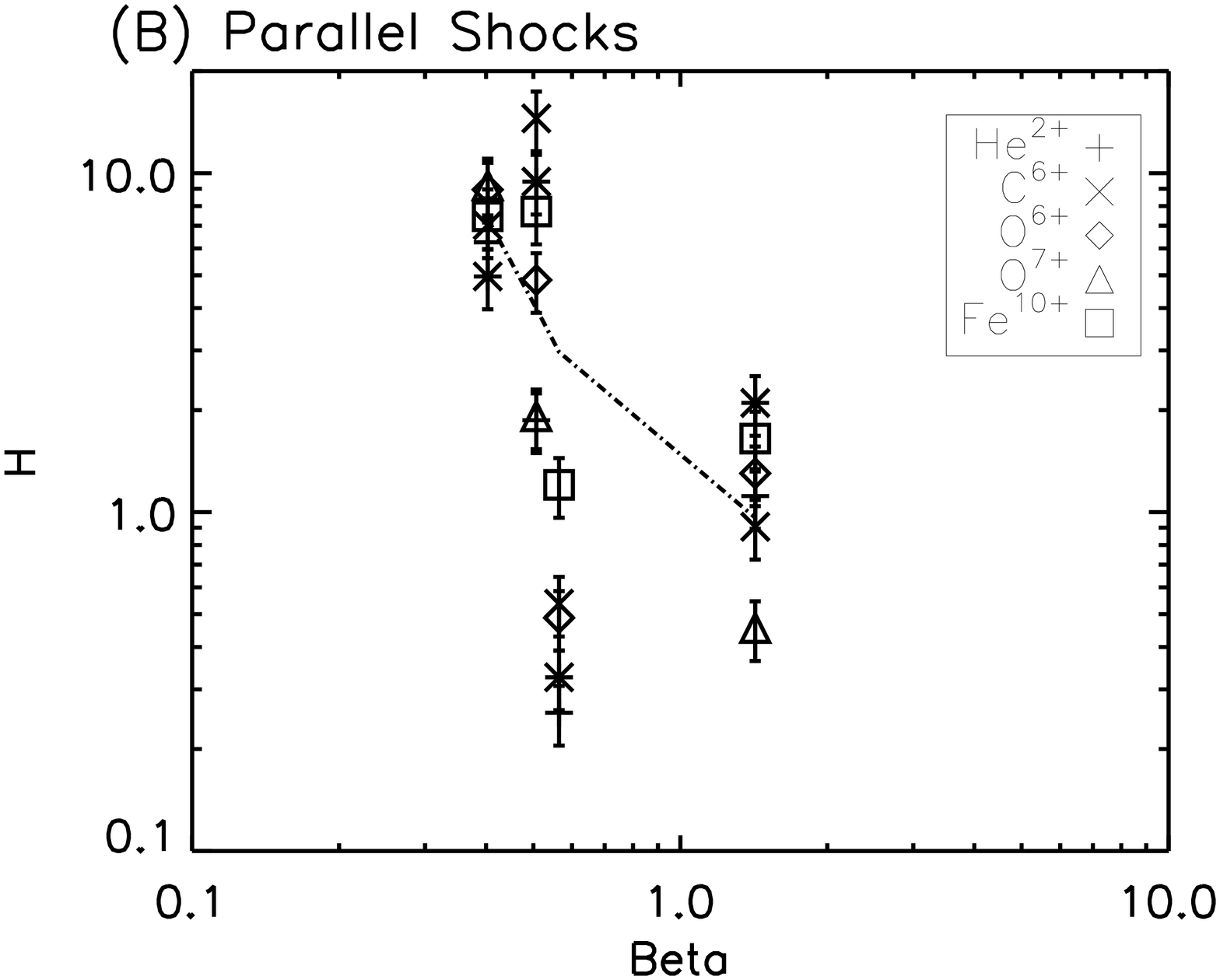}
\caption{\label{betaf} Plot of $\beta$ versus heating, H, for all the
heavy ions in a shock.  The heating is the ratio of the square of the
downstream ion thermal velocity to the square of the upstream ion
thermal velocity. $\beta$ is the ratio of thermal to magnetic energy
densities.  The line is the exponential fit to the data.}
\end{figure}

Plasma $\beta$ is plotted versus the heating for each ion species for
perpendicular shocks in Figure \ref{betaf}a.  The plot shows that with
increasing $\beta$ the heating of the ions decreases. The heating is
therefore more effective in magnetically dominated regimes.  The lines plotted are the fit to an exponential fit to the data.

For the parallel shocks, Figure \ref{betaf}b, there is an observed
decrease in heating with increasing $\beta$, however, only a few
shocks are available for study and hence the trend is more uncertain
but decreases more sharply than the perpendicular shocks.

\subsection{Heating and Mass to Charge Ratio}

\begin{figure}
\includegraphics[angle=90,clip=,scale=0.4]{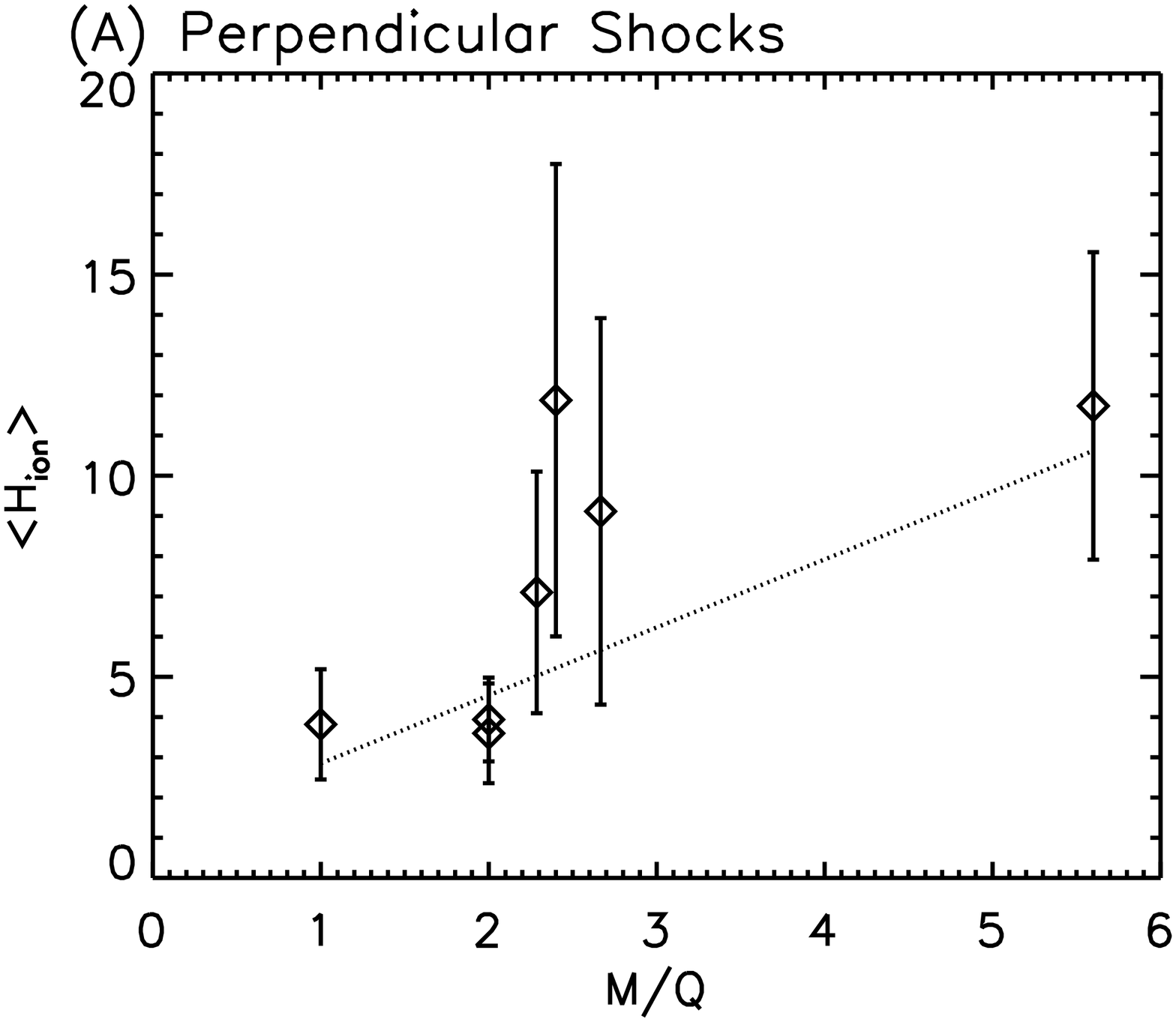}
\includegraphics[angle=90,clip=,scale=0.4]{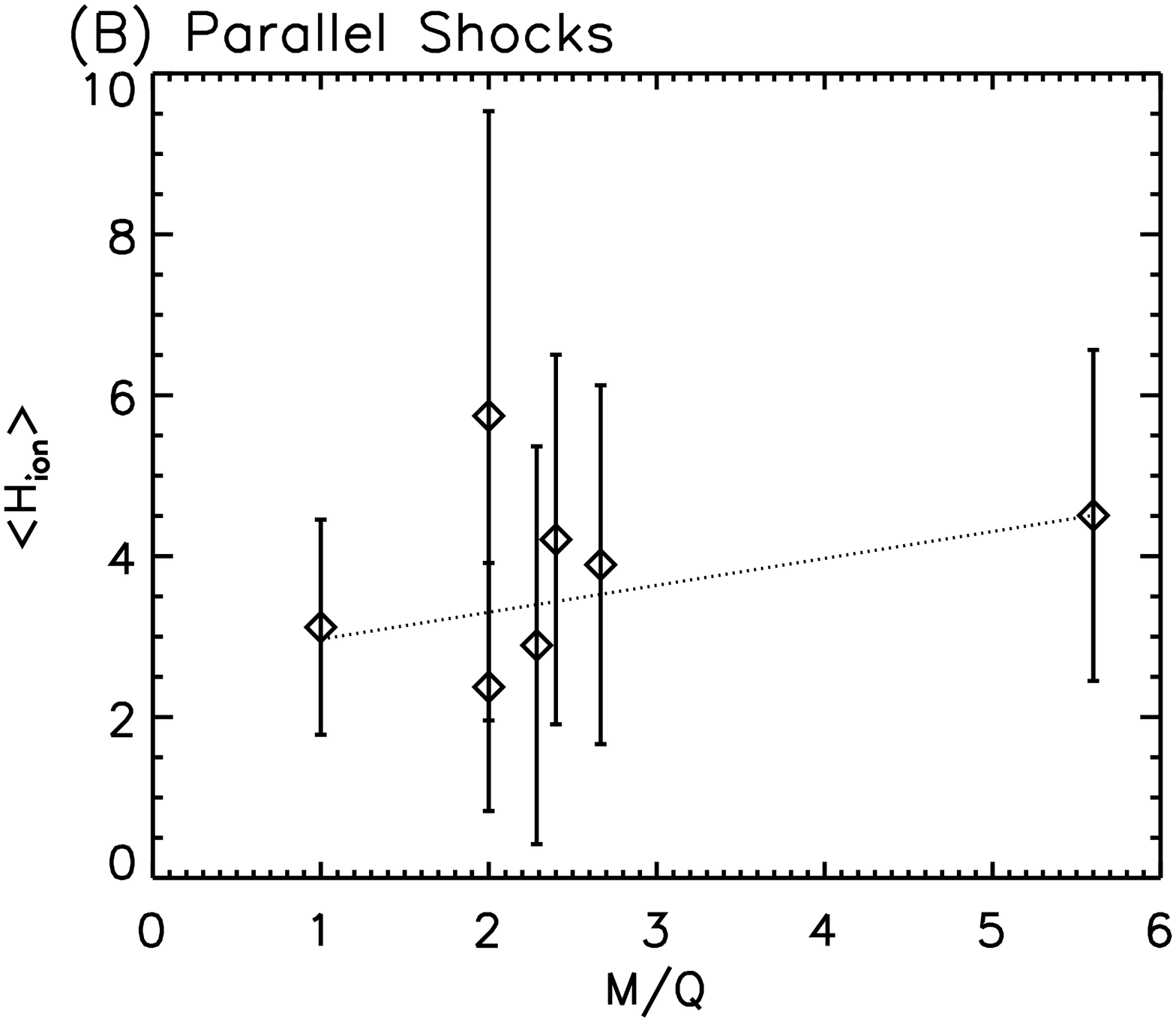}
\caption{\label{mqf}Plot of M/Q versus heating, H, for all the ions in
a shock.  The heating is the ratio of the square of the downstream ion
thermal velocity to the square of the upstream ion thermal velocity.
The heating for each ion was averaged in order to determine a trend in
the data.  The line is the least square fit to the data.}
\end{figure}

The mass to charge ratio of an ion determines the type of wave and
particle interaction that the ion can experience.  In Figures
\ref{mqf}a and \ref{mqf}b, the average heating per ion is plotted
versus the mass to charge ratio for perpendicular and parallel shocks
respectively.   The line is a least squared fit to the mean values
with the standard deviation of the mean values as error bars.  For the
perpendicular shock the fit has a reduced chi-squared value of
$\chi_{\nu}^{2}$=0.25, for the expression H=1.15+01.6M/Q.  The fit for
the parallel shocks, $\chi_{\nu}^{2}$ $\sim$0.48, the line is present
to emphasize the trend towards an increase in heating with increasing
M/Q ratio.

In the perpendicular shocks both the M/Q=2 ions are heated approximately the same, whereas in the parallel shocks they are heated differently, making the mass important in the parallel shock given that the larger value is the C$^{6+}$ ion, where the M/Q ratio is important for the perpendicular shock.

These shock exhibit an increase in heating based on a larger M/Q ratio
regardless of the exact $\theta_{Bn}$, however, the effect is more
pronounced in the perpendicular shock.  \citet{lee00} found that due to transmission of ions across the shock, the gyro speed, v$_{g}$, is approximately equal to the thermal speed.  The gyro speed increased as the M/Q ratio increased which agrees with this data set.

\section{Discussion of Heating Mechanisms}

A shock heating mechanism is needed that addresses all observational
constraints found in this study.  Any proposed mechanism must meet the
following criteria:

\begin{enumerate}
\item Heating increases with decreasing $\beta$
\item Perpendicular shocks heat more effectively than parallel shocks
\item The ions that initially have thermal velocities nearly equal to
protons are preferentially heated
\item Heating increases with increasing mass-to-charge ratio.
\item Increasing Mach Number increases the heating
\end{enumerate}

\citet{lee00} modeled heating in the solar corona.  This heating
mechanism and the predictions for the downstream thermal speed will be
used to evaluate this mechanism for the CME shock heating.  The protons are the main species and will form the shock while the heavy ions will act as test particles through the shock front.  The
simplifying assumption will be made that all changes in the ions
velocity will occur parallel to the shock normal.  The proton speed is also assumed to be the speed of the solar wind or the upstream bulk speed.  As protons pass
through the shock they are slowed according to the Rankine-Hugoniot
conditions in order to conserve energy.  Their bulk kinetic energy is
changed into potential energy forming an effective electrostatic
potential.  Therefore, the potential is proportional to the change in
energy of the proton.

\begin{equation}
\label{potential}
q\phi \sim \frac{1}{2}m_{p}v_{p,u}^{2}-\frac{1}{2}m_{p}v_{p,d}^{2}
\end{equation}

This potential jump is seen by the heavy ions as they approach the
shock front.  But the heavy ions conserve energy, increasing their
velocity across the shock according to their M/Q ratio.

\begin{equation}\label{ke}
\Delta
K.E._{i}=q_{i}\phi=\frac{1}{2}m_{i}(v_{i,u}^{2}-v_{i,d}^{2})\end{equation}

where q$_{i}$ is the charge of the ion, m$_{i}$ is the mass of the ion
in units of proton mass, v$_{i,u}$ is the velocity of the ion upstream
and v$_{i,d}$ is the velocity of the ion downstream of the shock.
Plugging in the potential formed by the protons (\ref{potential}) to
the kinetic energy of the heavy ions (\ref{ke}), we can find an
expression for the downstream ion velocity in terms of the M/Q
($\alpha$) and observed parameters.

\begin{equation}
\label{ionheating}
v_{i,d}=v_{p}\sqrt{\frac{(\alpha-1)+c^{2}}{\alpha}}
\end{equation}

where $\alpha$ is the mass to charge ratio, v$_{p}$ is the solar wind
proton velocity, and c is the ratio of downstream proton velocity to
upstream proton velocity in the shock frame.

The velocity computed in Equation \ref{ionheating} is the kinetic
velocity of the particle at the shock front.  However, we are trying
to examine the thermal speed of the heavy ions.  As the distribution
of the ions moves away from Maxwellian, the thermal speed becomes a
measure of kinetic energy.  \citet{lee00} derived that for a perpendicular shock the gyro velocity was approximately equal to the thermal velocity.  Using the definition of gyro velocity from the \citet{lee00} paper,
\begin{equation}
\label{vgyro}
v_{i,d,th}=v_{gyro}=v_{p,u}|\sqrt{(1-\frac{\alpha Q}{M})}- \frac{B_{t,1}}{B_{t,2}} |
\end{equation}

where $\alpha$ is defined as $e \Delta$$\phi$/$\epsilon$$_{0}$ - the ratio of electric potential at the shock front to the kinetic energy of protons upstream , v$_{p,u}$ is the upstream solar wind
proton velocity, B$_{t,1}$ is the upstream magnitude of the tangential magnetic field, and B$_{t,2}$ is the downstream magnitude of the tangential magnetic field.

By inspection, the downstream thermal velocity does increase with
increasing M/Q ratio.  Using the definition of a Mach number the
relationship between the upstream proton velocity and the Mach numer
is found to be v$_{p,u}$=M$_{A}$v$_{A}$.  Inserting this relation into
the square of Equation \ref{vgyro} in order to compare with the ion heating
data, the heating of the ions is proportional to the square of the
Mach number.  However, in the actual data sets, there is an increase
of heating with respect to increasing Mach number however not a
significant heating as suggested by this equation.

The plasma $\beta$ is inversely correlated with the square of the
Alfven velocity.  With respect to plasma $\beta$ the data match the
predicted trend of Equation \ref{vgyro} as the heating decreased with
increasing $\beta$.

As the Mach number increases we expect to have greater shock
compression of the magnetic field which would lead to a larger B$_{t,2}$.  The Mach number is not as strongly correlated as the magnetic characteristics, such as $\theta$$_{Bn}$ or $\beta$, with the heating of the heavy ions, however, the bulk motion provided by shocks at higher Mach numbers is necessary for the compression of downstream magnetic fields that influence the gyro velocity.

\section{Conclusions}

This study examined different ion species in the solar wind to
investigate the heating that occurs at a collisionless shock front
ahead of ICMEs. Based on the magnetic field angle to the shock normal,
$\theta$$_{Bn}$, each shock was analyzed for its dependence on Mach
number, M$_{A}$, plasma $\beta$, and the behavior of ions based on
their mass to charge ratio, M/Q. An increasing $\beta$ was found to
reduce the ion heating at the shock, highlighting the importance of
the magnetic field to the heating process. As the Mach number
increased the heating also increased.  As the mass to charge ratio
increases, the heating of the ion increases.  Quasi-perpendicular
shocks were shown to heat the ions more efficiently than the
quasi-parallel shocks.

A heating method based on a post-shock potential described by
\citet{lee00} was examined for the heating trends observed in this
study.  The data and the predicted values of heating were not well
correlated for the parallel shock but did match trends found for the
M/Q, M$_{A}$ and plasma $\beta$ trends for the perpendicular shocks.
The mechanism reveals an increased heating with increasing Mach
number.  However, the magnitude of the increase in heating with Mach number is not of the same order of magnitude that is found in the data and the trend however weak is inverse to that found by \citet{sch88}.

There is currently similar ion heating data available for
collisionless shocks in supernova remnants showing non-preferential
heating to ions.  Using the data compiled here we can compare the
heating ratios of ions to protons downstream and use that ratio to find
a plasma $\beta$ for the supernova remnant (SNR). In \citet{kor04}, it was
found that the ratio of oxygen to proton temperature post-shock was
8.3. Using the fit of heating versus plasma $\beta$ for the CME data, we obtain a $\beta$ for SN1006 of 1.7.  Knowing the
temperature, T=1.5 $\times$10$^{9}$K, and the density given, 0.25
cm$^{-3}$, we can also find a magnetic field of 1$\mu$Gauss.  This is
very close to the assumed galactic magnetic field of 3$\mu$Gauss. These
types of diagnostics could lead to a better understanding of the magnetic
field in SNRs and their role in ion heating and acceleration.

This begs the question of the ubiquitous nature of the shock physics:
what is the dominant factor to determine effect heating at a
collisionless shock front?  The supernova has a Mach number 10 times
that of the CME shocks however, as seen in this data set, the Mach
number is not the only factor in determining heating.
Density and magnetic energy seem to be of greater importance.

\acknowledgments
The ACE science center is found at \url{http://www.srl.caltech.edu/ACE/}. 
This work was performed with the support of ACE contract number.  K. Korreck acknowledges the Rackham-NSF Fellowship for funding this
work. This work made use of NASA's Astrophysical Data System.

Facilities: \facility{ACE}.

\clearpage

\clearpage

\clearpage
\clearpage

\clearpage
\clearpage

\end{document}